\documentclass[aip,reprint,amsmath,amssymb]{revtex4-1}

\usepackage{graphicx}
\usepackage{dcolumn}
\usepackage{float}
\usepackage{bm}
\usepackage{amssymb}
\usepackage{lipsum}
\usepackage[usenames,dvipsnames]{xcolor}

\begin{document}

\title{Entropic force of cone-tethered polymers interacting with a planar surface}

\author{James M. Polson}
\email{jpolson@upei.ca}
\affiliation{ Department of Physics, University of Prince Edward Island,
550 University Avenue, Charlottetown, Prince Edward Island, C1A 4P3, Canada }
\author{Roland G. MacLennan}
\affiliation{ Department of Physics, University of Prince Edward Island,
550 University Avenue, Charlottetown, Prince Edward Island, C1A 4P3, Canada }

\begin{abstract}
Computer simulations are used to characterize the entropic force of one or more
polymers tethered to the tip of a hard conical object that interact with a nearby hard flat
surface. Pruned-enriched-Rosenbluth-method (PERM) Monte Carlo simulations are used
to calculate the variation of the conformational free energy, $F$, of a hard-sphere polymer
with respect to cone-tip-to-surface distance, $h$, from which the variation of the entropic
force, $f\equiv |dF/dh|$, with $h$ is determined. We consider the following cases:
(1) a single freely-jointed tethered chain, (2) a single semiflexible tethered chain,
and (3) several freely-jointed chains of equal length each tethered to the cone tip.
The simulation results are used to test the validity of a prediction by Maghrebi 
{\it et al.} (EPL, {\bf 96}, 66002(2011); Phys. Rev. E {\bf 86}, 061801 (2012))
that $f\propto (\gamma_\infty-\gamma_0) h^{-1}$, where $\gamma_0$ and $\gamma_\infty$
are universal scaling exponents for the partition function of the tethered polymer
for $h=0$ and $h=\infty$, respectively. The measured functions $f(h)$ are generally
consistent with the predictions, with small quantitative discrepancies arising
from the approximations employed in the theory. In the case of multiple tethered
polymers, the entropic force per polymer is roughly constant, which is qualitatively
inconsistent with the predictions.
\end{abstract}

\maketitle

\section{Introduction}
\label{sec:intro}

Confinement of a polymer chain to a sufficiently small space distorts its shape,
leading to a significant reduction in its conformational entropy and thus an
increase in its free energy. The effects of confinement on single-polymer conformational
statistics has been the subject of numerous theoretical and computational studies, which
have examined variety of confinement geometries, including
cavities,\cite{cacciuto2006self, polson2015polymer, sakaue2018compressing, polson2019polymer}
channels,\cite{dai2016polymer, werner2017one, chen2018self, polson2017free, polson2018free} 
slits,\cite{dai2012systematic, nikoofard2014accuracy, tree2014odijk,
cheong2018evidence, teng2021statistical} 
as well as more complex geometries.\cite{klotz2015measuring, klotz2015correlated} 
Most relevant to the present work is the case
of a polymer confined to a slit between two parallel hard walls.\cite{nikoofard2014accuracy,
tree2014odijk, cheong2018evidence,teng2021statistical} These studies have characterized the
scaling  of the molecular dimensions and free energy with respect to confinement dimension,
contour length and persistence length. Distinct scaling regimes such as the de~Gennes, 
extended  de~Gennes and Odijk regimes have been identified, and some predictions have
been verified by experiments employing fluorescently labeled DNA molecules confined to
nanoslits.\cite{leith2016free}

\begin{figure}[!ht]
\begin{center}
\vspace*{0.2in}
\includegraphics[width=0.38\textwidth]{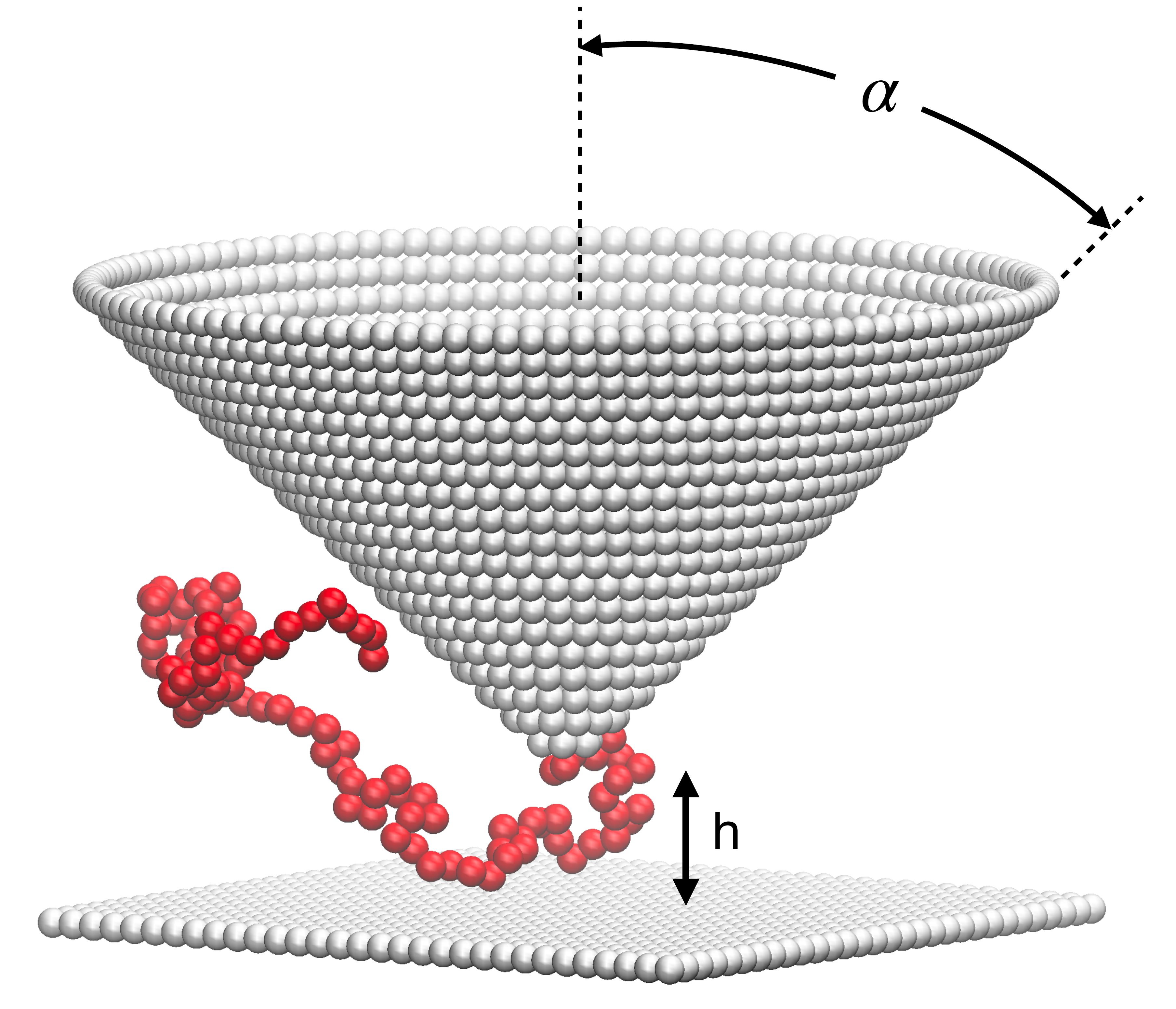}
\end{center}
\caption{Illustration of the system examined in this study. A polymer is tethered to the
tip of a hard conical object in the vicinity of hard planar surface. The cone and the
surface both extend to infinity. The cone half-angle, $\alpha$, and the cone-tip-to-surface
distance, $h$, are both illustrated. We examine cases of single flexible and semiflexible
chains as well as multiple flexible chains tethered to the cone tip.}
\label{fig:cone_illust}
\end{figure}

A system conceptually similar to that of a slit-confined polymer was examined
a number of years ago Maghrebi {\it et al.}.\cite{maghrebi2011entropic,maghrebi2012polymer}
In these studies, a polymer was confined to the space between a flat surface and a hard cone,
with one end of the polymer tethered to the tip of the cone. The system is 
illustrated in Fig.~\ref{fig:cone_illust}. Note that this reduces to
a slit-confined tethered polymer when the cone angle $\alpha$ (defined in
the figure) is $90^\circ$. As is the case for confinement between parallel walls, the
conformational free energy increases upon a reduction in the distance between the surfaces, 
here defined as the cone-tip-to-surface distance, $h$. Remarkably, the entropic force,
$f\equiv |dF/h|$, is expected to satisfy the simple universal relation,
$f = {\cal A} k_{\rm B}T/h$, where $k_{\rm B}$ is Boltzmann's constant and $T$ is absolute
temperature. The relation is expected to hold for sufficiently long chains in the regime 
where $a\ll h \ll R_{\rm g}$, where $a$ is the link length (monomer size or persistence 
length)  and $R_{\rm g}$ is the root-mean-square radius of gyration of a free polymer. 
The proportionality constant,  ${\cal A}$,  depends solely on  basic geometrical factors 
and gross features of the  polymer and is of the order of unity. 
The scaling relation also holds more generally for other systems provided the two obstructing
surfaces are scale invariant. (Other examples include pyramids and wedges.)
Maghrebi {\it et al.} calculated ${\cal A}$ for both ideal and real (i.e., self-avoiding) 
polymers  using analytical, simulation, and $\epsilon$-expansion methods.
The $\epsilon$-expansion was also used to estimate the effects of multiple polymers
tethered to the cone tip on the entropic force.

The theoretical analysis of the tethered-polymer system examined in
Refs.~\onlinecite{maghrebi2011entropic} and \onlinecite{maghrebi2012polymer} was
intended to provide motivation for future experimental measurements of entropic forces of
polymers using atomic force  microscopy (AFM). In this scenario, the cone represents 
the AFM cantilever tip, at the end of which is attached one or more polymers. 
As the prefactor ${\cal A}$ is expected to be of order unity, the
measured force for a single tethered polymer at room temperature is predicted to be of 
order 0.1~pN for a distance of order 0.1~$\mu$m, which they noted was just at the margins 
of measurement for precision AFM devices. On the other hand, the authors also noted that
using multiple polymers tethered to the AFM tip would increase the magnitude of
the force without changing the form of the force-distance relation. Recently, such
experiments were carried out by Liu~{\it et al.}, who measured the  entropic force of 
polyethylene glycol (PEG) polymers tethered to pyramidal  cantilever tip in the vicinity 
of a flat surface.\cite{liu2019measurement} The measurements were made in salt water
solution to minimize the effects of electrostatic forces and a hydrophobic plate to 
minimize polymer adhesion forces. Force-distance curves were measured and compared to 
the prediction of Maghrebi {\it et al.}, as well as to the Alexander-de~Gennes (AdG) 
theory for a polymer brush.\cite{deGennes_book} Each analysis suggested that a few tens of
polymers had been tethered to the cantilever. Notably, the AdG theory appeared to provide
a more accurate prediction, calling into question either the relevance of the theory of
Maghrebi {\it et al.} to such experiments or else the appropriateness of the design of
this particular experiment to test the theory. 

The purpose of the present study is to use computer simulations to examine the accuracy
of the prediction for the force-distance relation, $f={\cal A}k_{\rm B}T/h$. As noted
above, the relation is expected to hold only for polymers that are sufficiently long
and only over a restricted range of $h$ where that distance is the only relevant length
scale. Systems for which these conditions are only marginally satisfied are expected to show
deviations from the predictions, and a key goal of the study is to quantify such effects. 
The calculations can be used to determine the origin of the discrepancy between theory
and the experiments of Ref.~\onlinecite{liu2019measurement} and to provide insight for use
in future experiments that can be designed to provide a more meaningful test of the theory. 
The other goal of the study is more fundamental. In their first study on this 
topic,\cite{maghrebi2011entropic} Maghrebi {\it et al.} write: ``The simple force law... 
follows easily from various polymer scaling forms... and should be part of polymer lore.
Surprisingly, we could not find an explicit reference to it in any of the standard polymer
textbooks.'' Over recent decades, computer simulation methods have provided valuable
insight into the many other scaling predictions in polymer physics that do appear in the
standard texts, and the present work should help address the omission noted by the authors.

The remainder of the article is organized as follows. In Sec.~\ref{sec:theory}, we
review the derivation of the prediction for the force-distance relation, highlighting the
various approximations that are employed. In Sec.~\ref{sec:model}, we briefly describe
the simple molecular model used in the simulations. Section~\ref{sec:methods} outlines
the MC simulation methods that were employed for the calculations. Section~\ref{sec:results}
presents the results for three variations of the model polymer: (1) a single tethered 
flexible chain, (2) a single tethered semiflexible polymer, and (3) multiple tethered
flexible chains. Finally, in Sec.~\ref{sec:conclusions} we summarize the key findings
of the study.

\section{Theory}
\label{sec:theory}

In this section we provide a brief review of the derivation of the force-distance
relation for the cone-tethered polymer system first presented by Maghrebi 
{\it et al.}, \cite{maghrebi2011entropic, maghrebi2012polymer} highlighting the
various approximations that are employed in the process.

Consider a single self-avoiding polymer chain tethered to the apex of a 
hard cone of half-angle $\alpha$. The tip of the cone, and thus the tethered monomer, is 
located a distance $h$ from an infinite, hard flat surface, whose normal is aligned with 
the symmetry axis of the cone. The system is illustrated in Fig.~\ref{fig:cone_illust}.
The special cases of $h=0$ and $h=\infty$ correspond to scale-free confinement geometries. 
(Note: In this article, $h=\infty$ is used to denote the case of no polymer-plane
interactions for polymers of arbitrary length.) In such cases, the partition function 
for a self-avoiding polymer has the form\cite{deGennes_book}
\begin{eqnarray}
Z = b q^N N^{\gamma-1},
\end{eqnarray}
where $q$ is the effective coordination number of the polymer, $b$ is a model dependent
coefficient of  order unity whose value depends on $h$ (i.e., $h=0$ or $h=\infty$), and 
$\gamma$ is a universal exponent  that depends on the cone angle $\alpha$. 
Since the conformational free energy of the polymer is given by  $F/k_{\rm B}T = -\ln Z$, 
the free energy difference between the systems for $h=0$ and  $h=\infty$ is given by
\begin{eqnarray}
\Delta F/k_{\rm B}T = C + \Delta\gamma\ln N,
\label{eq:FCDg}
\end{eqnarray}
where $\Delta F \equiv F_0 - F_\infty$, $C\equiv -\ln(b_0/b_\infty)$ and $\Delta \gamma
\equiv \gamma_\infty-\gamma_0$, and where the subscripts refer to the cases of $h$=0 and 
$h$=$\infty$.

Now consider the case of arbitrary tip-to-surface distance $h$. As $h$ decreases and the cone
comes closer to the flat surface, the polymer becomes increasingly confined, and thus its 
conformational entropy decreases. This in turn effects an increase in the free energy, $F$. 
The magnitude of the entropic force, defined as $f \equiv \left| dF/dh \right|$, is also
expected to increase with a reduction in $h$.  Using dimensional analysis, Maghrebi 
{\it et al.}  have argued that the entropic force  should vary  inversely with $h$,
\begin{eqnarray}
f = {\cal A} \frac{k_{\rm B}T}{h},
\label{eq:fdef}
\end{eqnarray}
in the regime where $a\ll h \ll R_{\rm g}$, where $a$ is the size of a link in the polymer chain
(i.e. the monomer width for a freely-jointed chain model and the persistence length in the case
of a semiflexible polymer), and $R_{\rm g}$ is the root-mean-square radius of gyration of a 
free polymer. This scaling ansatz follows from the fact that $h$ is the only relevant length
scale in this regime. The quantity ${\cal A}$ depends only on geometric factors, such as the
cone angle. They estimate ${\cal A}$ as follows. The work done against the entropic force in
bringing the cone  from far away to make contact with the plate is calculated:
\begin{eqnarray}
W = \int_{a}^{R_{\rm g}} dh f(h) = {\cal A} k_{\rm B}T \ln\left(\frac{R_{\rm g}}{a}\right).
\label{eq:W}
\end{eqnarray}
The upper bound on the integral arises from the fact that $f\approx 0$ until 
$h\approx R_{\rm g}$  and the lower bound is due to the fact that $f$ is too large for the 
cone to approach the surface at  distances below $a$. The radius of gyration scales as 
$R_{\rm g}/a = c N^\nu$, where $N$ is the number of polymer links and where $\nu\approx 0.588$ 
for a self-avoiding polymer.\cite{Rubinstein_book} The scaling prefactor, $c$, is of order 
unity and depends on the details of the
molecular model. Substitution into Eq.~(\ref{eq:W}) yields:
$W = {\cal A} \nu k_{\rm B}T \ln(N) + {\cal A} \nu k_{\rm B}T\ln(c)$.
Since $c$ is of order unity, we can neglect the second term, which gives
\begin{eqnarray}
W = {\cal A} \nu k_{\rm B}T \ln(N).
\label{eq:W2}
\end{eqnarray}
The work done against the entropic force is simply equal to $\Delta F$, the change in the 
free energy in moving the cone from far away to a point where it is in contact with the 
surface. Comparing Eq.~(\ref{eq:FCDg}) (but ignoring the additive constant $C$) and
Eq.~(\ref{eq:W2}), it follows: ${\cal A} = \Delta\gamma/\nu$, where $\Delta\gamma\equiv
\gamma_\infty-\gamma_0$. Consequently,
\begin{eqnarray}
f = \frac{\Delta\gamma}{\nu}\frac{k_{\rm B}T}{h}.
\label{eq:ftheory}
\end{eqnarray}
Equation~(\ref{eq:ftheory}) thus predicts that the entropic force scales inversely
with the tip-to-surface distance $h$. The proportionality factor $\Delta\gamma$ 
depends on the cone angle $\alpha$. 
As $\alpha$ increases, the degree of confinement 
also increases.  This will result in a reduction in conformational entropy and an
expected increase the entropic force. Thus, $\Delta\gamma$ is expected to increase
monotonically with increasing $\alpha$. Note that $f$ is independent of the polymer 
length $N$ as well as the persistence length. Thus, the functions $f(h)$ for 
different values of $N$ and $\kappa$ are expected to overlap in the regime where the condition 
$a\ll h \ll R_{\rm g}(N)$  is satisfied for each polymer chain. A key goal of the 
present study is to test the validity of  the prediction in Eq.~(\ref{eq:ftheory}). 

\section{Model}
\label{sec:model}

We employ a very simple, athermal model in our simulations.
The model consists of one or more polymer chains tethered
to the tip of a hard, conical object in the vicinity of a hard, flat
surface whose normal is parallel to the symmetry axis of the cone. 
In all cases, the polymer is modeled as a chain of $N+1$ hard spheres, with sphere
diameter and fixed bond length both equal to $\sigma$, which defines the
length scale. We consider the cases of both a freely-jointed chain and
a semiflexible chain. In the latter case, the bending rigidity of the polymer 
is modeled using a bending potential with the form, $u_{\rm bend}(\theta)=
\kappa(1 - \cos\theta)$. The angle $\theta$ is defined for a consecutive
triplet of monomers centered at monomer $i$ such that 
$\cos\theta_{i}=\hat{u}_{i}\cdot\hat{u}_{i+1}$,
where $\hat{u}_{i}$ is the unit vector pointing from monomer $i-1$ to monomer $i$. 
The bending constant $\kappa$ determines the overall stiffness of the polymer and is 
related to the persistence length $P$ by\cite{micheletti2011polymers} 
$\exp(-\langle l_{\rm bond} \rangle/P) = \coth(\kappa/k_{\rm B}T) - k_{\rm B}T/\kappa$,
where $\langle l_{\rm bond}\rangle$ is the mean bond length.
For our model, the bond length is fixed to $l_{\rm bond}=\sigma$. 
Note that for sufficiently large $\kappa/k_{\rm B}T\gg 1$ this implies $P/\sigma\approx
\kappa/k_{\rm B}T$.

The polymer is tethered to the tip of a hard conical structure of half-angle $\alpha$.
The center of the first monomer is located at the exact tip of the cone. The 
cone-tip-to-surface distance $h$ is defined such that $h=0$ corresponds to the
tethered monomer being in contact with the surface. Configurations in which monomers 
overlap with each other, the cone, or with the surface have an energy of infinity
and thus are forbidden. For the case of multiple tethered polymers, each polymer shares
the same end monomer that is tethered to the cone tip. All simulations for multiple-polymer
systems used only the freely-jointed chain model.

\section{Methods}
\label{sec:methods}

We use Pruned-enriched Rosenbluth method (PERM) simulations to 
calculate the excess free energy of the tethered polymer. 
PERM is a chain growth MC method that uses a dynamic bias to obtain
importance sampling. PERM is based on the Rosenbluth-Rosenbluth (RR)
method, which can also be used to calculate polymer free energies. PERM uses
a more sophisticated algorithm that enables it to overcome the well
known attrition problem that limits the applicability of the RR method to
rather short polymers of ${\cal O}(10^2)$ segments. By contrast, PERM can
be used to grow polymer chains orders of magnitude greater in length. 
In this algorithm, a tree of chains, called a ``tour'', is grown using 
the operations of ``pruning'' and ``enriching'' in such a way as to 
dramatically reduce the attrition rate of the RR method.

We employ an off-lattice version of the algorithm developed by 
Tree {\it et al.}\cite{tree2013dna} 
The initial monomer is placed at the tip
of the cone and for the $n$th growth step a set of $K$ trial steps are calculated.
For freely-jointed chains, the orientation of the trial monomer position is chosen
with equal probability for all directions. For semiflexible chains, the orientations
are drawn from a probability distribution governed by the bending potential, 
$u_{\rm bend}$. Each trial step is assigned a Rosenbluth weight
\begin{eqnarray}
{a_n^{(k)} = \exp(-U_n^{(k)}/k_{\rm B}T).}
\end{eqnarray}
{Here, $U_n^{(k)}$ is the potential energy associated with the $k$th 
trial placement of monomer $n$. It includes non-bonded interactions with previously 
grown monomers (i.e. those with index $<n$), as well as monomer-wall interactions.  
Note that these interactions are athermal in character, i.e., the energy is infinity if a monomer 
overlaps with another monomer or with a confining wall and is zero otherwise.  Thus, 
$a_n^{(k)}=1$ if the trial move does not result in overlap and $a_n^{(k)}=0$ if it
does result in overlap.}  The weight of the $n$th growth step is defined
\begin{eqnarray}
w_n = \sum_{k=1}^{K} a_n^{(k)}. 
\end{eqnarray}
{Clearly, $w_n$ is an integer in the range $w\in[0,K]$ and is a
count of the number of trial moves that do not result in overlap.  }
To make one step, one of the trial steps is randomly chosen according to the
probability
\begin{eqnarray}
p_n^{(k)} = a_n^{(k)} / w_n.
\end{eqnarray}
The cumulative weight of the $n$th chain step is defined:
\begin{eqnarray}
W_n = \prod_{i=0}^{n} w_i.
\end{eqnarray}
This is an approximate count of the number of configurations generated using $K$
trial moves per step. 

In principle, the average of $W_n$ can be used to calculate the excess free 
energy of the polymer, $F_{\rm ex}(n)$,
\begin{eqnarray}
\beta F_{\rm ex}(n) = -\ln\left(\frac{Z_n}{Z_{{\rm id},n}}\right)
\approx -\ln\langle W_n \rangle
\label{eq:FW}
\end{eqnarray}
where
\begin{eqnarray}
F_{\rm ex}(N,h;\alpha) \equiv F(N,h;\alpha) - F_{\rm id}(N),
\label{eq:Fexdef}
\end{eqnarray}
where $\beta F=-\ln Z$ is the total conformational free energy of the polymer chain 
with configurational partition function $Z$, and where $\beta F_{\rm id}=-\ln Z_{\rm id}$ 
is the free energy of an ideal chain (i.e., in the absence of all monomer-monomer and
monomer-wall interactions) with partition function $Z_{\rm id}$.  While both $F$ and 
$F_{\rm id}$ depend on  chain length $N$, $F$ (and therefore $F_{\rm ex}$) also depends 
on the parameters associated  with the confining geometry, $h$ and $\alpha$. As the chain
grows, $W_n$ fluctuates and very quickly can approach zero if no trial positions can
be found that do not lead to monomer-monomer or monomer-wall overlap, thus leading
to the problem of attrition mentioned above. To overcome this problem, PERM uses
the procedures of pruning and enrichment to bias the chain growth toward successful
states, i.e., those without any overlap. In cases where $W_n$ rises above its ensemble
average $\langle W_n\rangle$, chain growth is taken to be successful, and the tour is
``enriched'' by spawning branches, or copies. If $W_n/\langle W_n\rangle$ falls then
chain growth  is struggling, and the tour is terminated along this branch, i.e. it is
``pruned''. The pruning rate and the number of copies created during enrichment 
are determined by the ratio, $W_n/\langle W_n\rangle$, using the stochastic,
parameterless procedure described by Prellberg and Krawczyk.\cite{prellberg2004flat}.
The continuous pruning and regrowing of the chain leads to a depth-first type
of diffusion along the chain contour length.\cite{grassberger1997pruned}

In Eq.~(\ref{eq:FW}), we note that the desired quantity, $F_{\rm ex}(n)$ depends on 
the average $\langle W_n \rangle$, which itself is used in the criterion for choosing
to prune or enrich the chain during execution of the growth process in the simulation. 
While $\langle W_n \rangle$ can be estimated during run-time and used to calculate
a more accurate estimate in a self-consistent calculation, initial estimates can be
poor, leading to a slow execution, particularly for longer chains. To overcome this
problem, we conduct a sequence of short runs to find an approximate estimate 
for $\langle W_n \rangle$. We then carry out a single long run using the estimate
to determine the pruning and enrichment probabilities in order to determine a much
more accurate value of $\langle W_n \rangle$ and, therefore, $F_{\rm ex}$. These
short runs are carried out as follows. We first grow a chain of length $N_{\rm inc}$
monomers using the RR method and calculate the $\langle W_n\rangle$ for 
$n\in [1,N_{\rm inc}]$. 
Then we grow a chain of length $2N_{\rm inc}$, using PERM with the previous estimate
of $\langle W_n\rangle$ to grow the first $N_{\rm inc}$ monomers and the RR method
to grow the next $N_{\rm inc}$ monomers. This yields an estimate of $\langle W_n\rangle$
in the range $n\in [0,2N_{\rm inc}]$. This process is applied iteratively for 
$N/N_{\rm inc}$ steps until a chain of length $N$ is grown and $\langle W_n\rangle$
is estimated for the full range of chain lengths. In a typical run for a polymer of
length $N$=5000 using $K=10$ trial steps used $N_{\rm inc}$=100 and thus 
$N/N_{\rm inc}$=50 increments with each simulation using $10^5$ tours, and the final 
run used $10^6$ tours. To maximize computational efficiency, we employ the neighbor-list
method described in the Supporting Information for Ref.~\onlinecite{tree2013dna}.

A system with multiple polymers tethered to the cone tip can also be thought
of as a star polymer with the same number of ``arms'' and with the branch point 
tethered to the tip. Modification of 
the algorithm for use with star polymers is straightforward and has been described
previously.\cite{hsu2004scaling, hsu2004effective, hsu2011review}
The most important detail is that each arm of the polymer is essentially grown
together simultaneously. To illustrate, consider a star polymer with arms that are
currently all of 
the same length. One monomer is added to the first arm, the next monomer is added to the 
second arm, and so on, until all arms are again of the same length, following
which the cycle begins again. In this study, we consider systems of up to 
$n_{\rm arm}=5$ arms each of length up to $N=1000$ monomers.

PERM simulations were used to calculate the excess free energy, 
$F_{\rm ex}(N,h;\alpha)$, for single tethered polymers up to a length of $N$=5000 
segments for cone-tip-to-surface distances in the range $h\in [0,300\sigma$] in integer
increments of $\sigma$ for freely-jointed chains and $h\in[0,400\sigma]$ for semiflexible 
chains. For star-polymer (i.e., multiple-chain) systems, we examined up to $n_{\rm arm}=5$ 
arms, each of length up to $N=1000$ monomers.  In addition, we use cone angles in the range 
$\alpha\in [10^\circ,90^\circ].$ The entropic force, $f\equiv |dF/dh| = |dF_{\rm ex}/dh|$ 
was calculated by first fitting $F_{\rm ex}(h)$ to a function typically of the form, 
$F=\exp(\sum_{n=0}^5 a_n (\ln(x+2))^n) + a_6$, by adjustment of the parameters $\{a_n\}$. 
This provided an excellent fit of the function over the full range of $h$. Subsequently,
$f(h)$ was determined from an analytical derivative of the fitting function.
{Typically, $f$ exhibited power-law scaling with respect to $h$ over an 
intermediate range of $h$. A fit of the calculated $f(h)$ in this range yields
an estimate of the scaling exponent.  In cases where the uncertainty in the scaling 
exponent was desired, a second analysis method was employed. First, a simple 
finite-difference method was used to estimate the derivative $dF/dh$ over this range.
A fit of these data then yielded an estimate of the scaling exponent,
which was typically very close to that obtained using the first method, as
well as an estimate of the uncertainty.}

In the results presented below, distances are measured in units of the monomer diameter,
$\sigma$, and energy is measured in units of $k_{\rm B}T$.

\section{Results}
\label{sec:results}

\subsection{A single tethered freely-jointed chain}
\label{subsec:freely}

PERM simulations were used to calculate the excess free energy, defined in Eq.~(\ref{eq:Fexdef}),
of a single self-avoiding freely-jointed hard-sphere chain tethered to the tip of a hard cone
located a distance $h$ from a hard flat surface. We define the free energy difference 
\begin{eqnarray}
\Delta F & \equiv & F(N,0;\alpha) - F(N,\infty;\alpha) \nonumber  \\
         & = & F_{\rm ex}(N,0;\alpha) - F_{\rm ex}(N,\infty;\alpha)
\label{eq:DelF}
\end{eqnarray}
where $h=0$ corresponds to the tethered monomer in contact with the surface and $h=\infty$ effectively
corresponds to a simulation in the absence of the flat surface. Thus, the difference in the
excess free energy measured in the two simulations with $h=0$ and $h=\infty$ yields the
difference in the total conformational free energy that appears in Eq.~(\ref{eq:FCDg}). 
Consequently, the measured difference in the excess free energy obtained from the two
PERM simulations should scale linearly with $\ln(N)$, with a proportionality constant
of $\Delta\gamma\equiv \gamma_\infty-\gamma_0$. 

\begin{figure}[!ht]
\begin{center}
\vspace*{0.2in}
\includegraphics[width=0.45\textwidth]{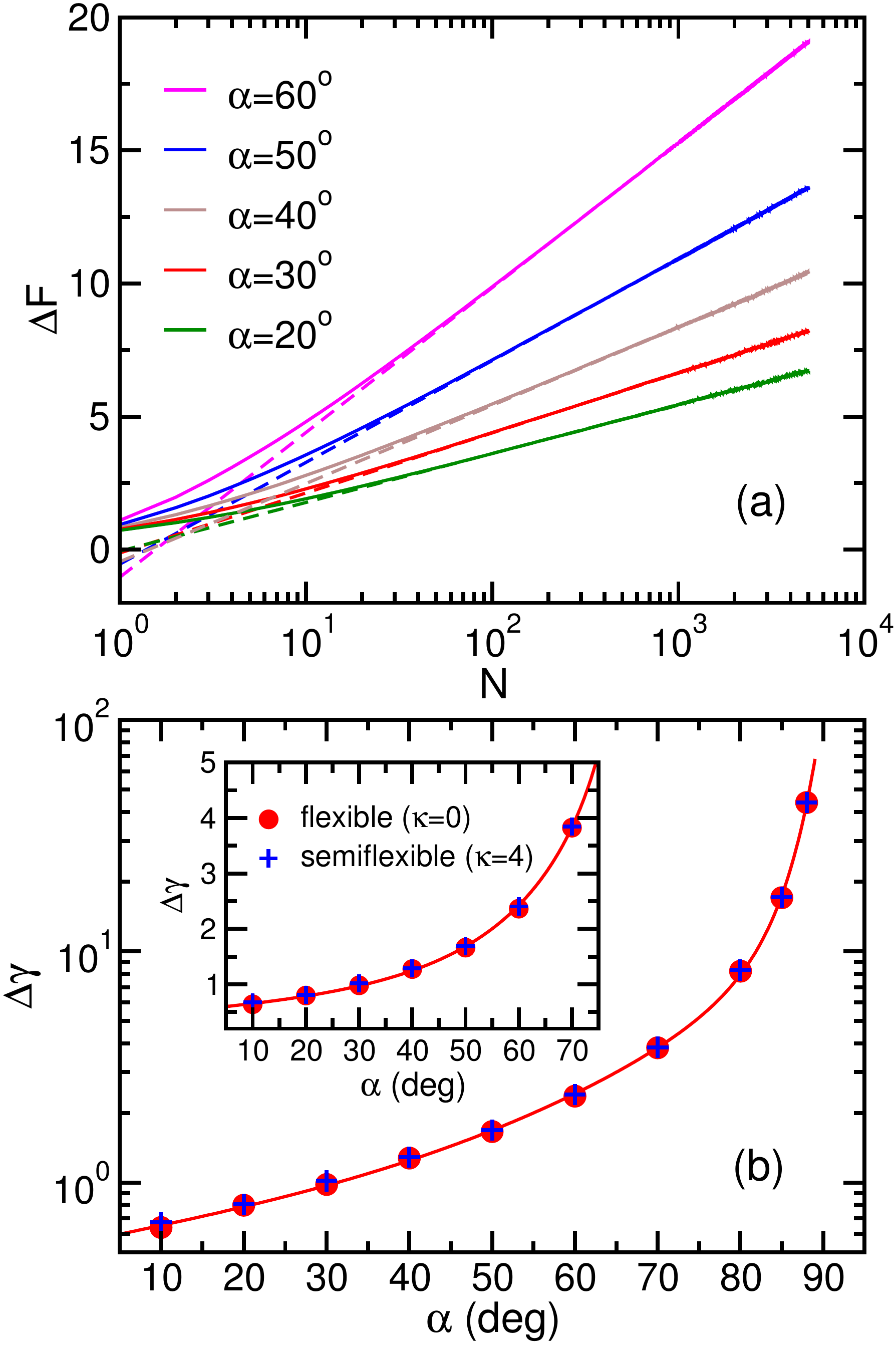}
\end{center}
\caption{(a) Free energy difference $\Delta F \equiv F(h=0)-F(h=\infty)$ vs polymer
length $N$ for a single freely-jointed chain tethered to a cone of half-angle $\alpha$.
Results for various values of $\alpha$ are shown. 
The dashed lines show fits of the data in the region $N\in [150,5000]$ to the function
$\Delta F = a_0 + a_1\ln N$. The fitted curves extend to lower values of $N$ to highlight
the discrepancy in this region. (b) Variation of $\Delta\gamma$ with $\alpha$, where
$\Delta\gamma\equiv \gamma_\infty - \gamma_0$ was obtained from the fits of the data
in panel~(a). The inset shows a close up of the data for low $\alpha$ plotted using
a linear scale for $\Delta \gamma$. The solid line is a guide for the eye. Overlaid
on these results are data for semiflexible chains with $\kappa$=4 for calculations
described in Sec.~\ref{subsec:semiflexible}. {Note that the uncertainties 
in the data points are smaller than the size of the symbols.}}
\label{fig:delF.N}
\end{figure}

Figure~\ref{fig:delF.N}(a) shows the variation of $\Delta F$ with $N$ for systems with
several values of $\alpha$. As expected, $\Delta F$ varies linearly with $\ln(N)$ for
sufficiently long chains. The dashed lines show fits to the data in the range 
$N\in [150,5000]$. The fitted curves are extended to the range $N<150$ to highlight
the expected discrepancy between the fit and the free energy in the low-$N$ regime.
The fits yield values of $\Delta\gamma$ for each cone angle, and the results are
plotted in Fig.~\ref{fig:delF.N}(b). As expected, $\Delta\gamma$ increases monotonically
with increasing cone angle $\alpha$ as a result of increasing confinement and a corresponding
loss of conformational entropy. The variation of $\Delta\gamma$ with $\alpha$ appears to be
consistent with that measured by Maghrebi {\it et al.}\cite{maghrebi2011entropic, maghrebi2012polymer}
In that study the excess free energy was calculated using a different method and for 
a lattice-model polymer system. In the limit of very large $N$ the two results
are expected to converge.

\begin{figure}[!ht]
\begin{center}
\vspace*{0.2in}
\includegraphics[width=0.45\textwidth]{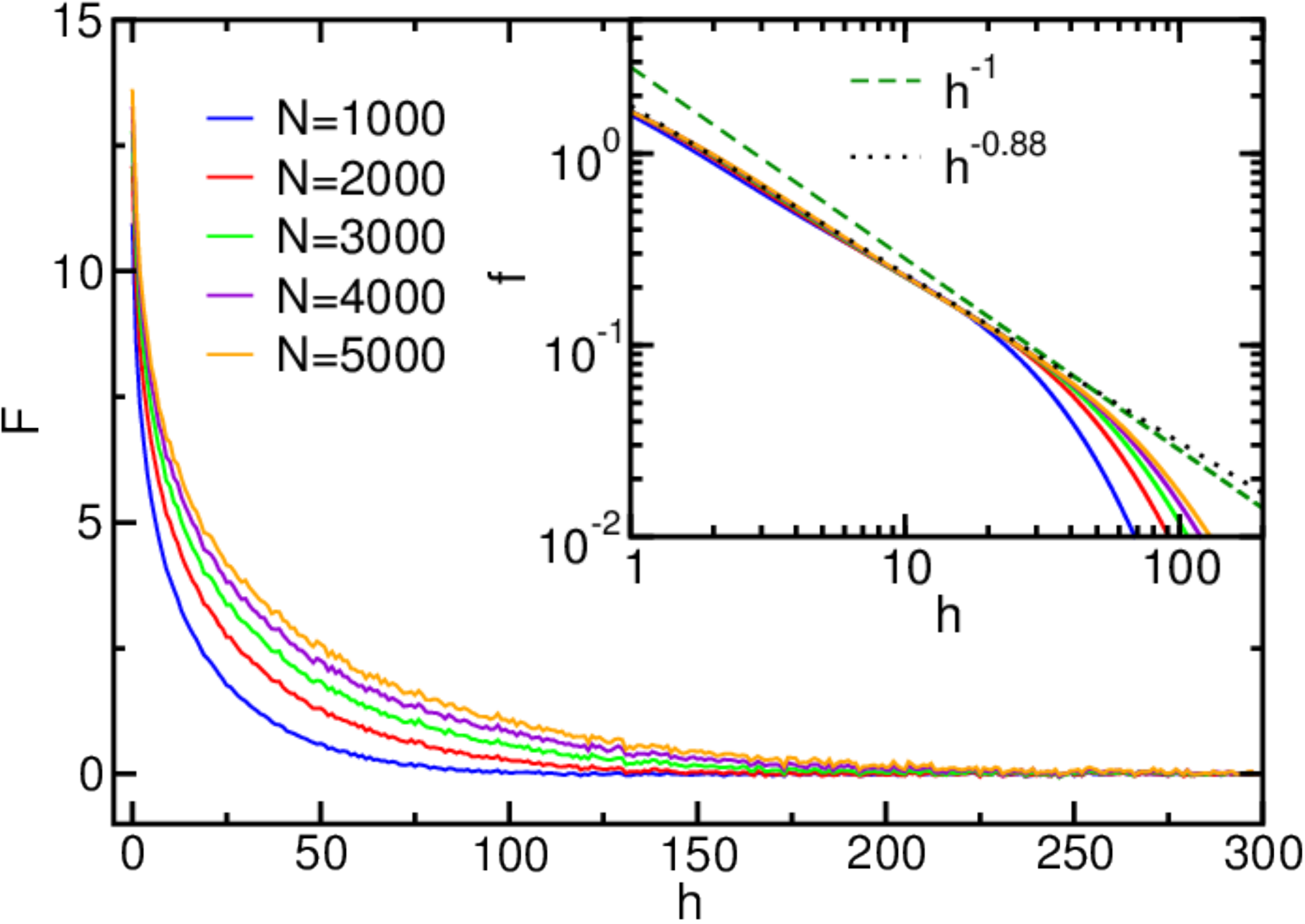}
\end{center}
\caption{ Free energy $F$ vs tip-to-surface distance $h$ for a cone angle of
$\alpha=50^\circ$. Results for various polymer lengths are shown.
The inset shows the variation of the entropic force, $f\equiv |dF/dh|$, with $h$.
The green dashed line shows the theoretical prediction using the value of
$\Delta\gamma$ for $\alpha=50^\circ$ in Fig.~\ref{fig:delF.N}(b). The dotted
black line shows a fit of the $N=5000$ curve to the function $f=cN^\mu$
in the region $h\in [4,20]$.  }
\label{fig:F.alpha50}
\end{figure}

Next, we consider the variation of the free energy, $F$, with the distance $h$. 
Note from Eq.~(\ref{eq:Fexdef}) that $F(h)$ differs from $F_{\rm ex}$, the quantity
actually calculated in the simulations, by a constant amount $F_0$, the free energy
of a free, ideal polymer. It is convenient to redefine $F(h)$, such that $F\rightarrow 0$
in the limit where $h$ is large, effectively by adding another constant.
Figure~\ref{fig:F.alpha50} shows the variation of $F$ with $h$ for a cone angle
of $\alpha=50^\circ$ and for various polymer lengths. The free energy was calculated
for tip-to-surface distances in the range $h=0-300$, where the upper bound was
chosen to be sufficiently large that $F$ has decayed to its large-$h$ limit 
by this point. The inset shows the corresponding variation of the  entropic force with 
$h$. The curves overlap at small $h$, but eventually diverge as $h$ increases. 
In the overlap region, $f$ obeys a power law. As $h$ increases, each curve
eventually peals away from this power-law curve, with shorter chains diverging before 
longer ones. The divergence arises from the violation of the
condition that $h\ll R_{\rm g}$ required for the validity of Eq.~(\ref{eq:ftheory}),
which predicts $f(h)$ to be independent of $N$. Shorter chains violate the condition 
before longer ones upon increasing $h$. Somewhat surprisingly, the curves largely overlap
and maintain the same power-law scaling right down to $h=1$, thus violating
the previously stated condition that $a\ll h$ (here, $a$=1 is the monomer size)
is required for the validity  of the theoretical prediction. Presumably, this
condition is of lesser importance than $h\ll R_{\rm g}$.

{To estimate the scaling exponent for $N=5000$ a finite-difference method 
was used to estimate the force in the range $h\in [4,30]$, and a subsequent fit 
to a power-law function yields scaling exponent of $-0.88\pm 0.08$.}
The fitted curve is qualitatively consistent with the theoretical curve calculated using
the value of $\Delta\gamma$ obtained from Fig.~\ref{fig:delF.N}. However, the quantitative
discrepancy is notable. The theory predicts an entropic force that is somewhat larger than
the measured values. In addition, the scaling exponent is close to, but somewhat smaller 
than the predicted value of $-1$. 

\begin{figure}[!ht]
\begin{center}
\vspace*{0.2in}
\includegraphics[width=0.45\textwidth]{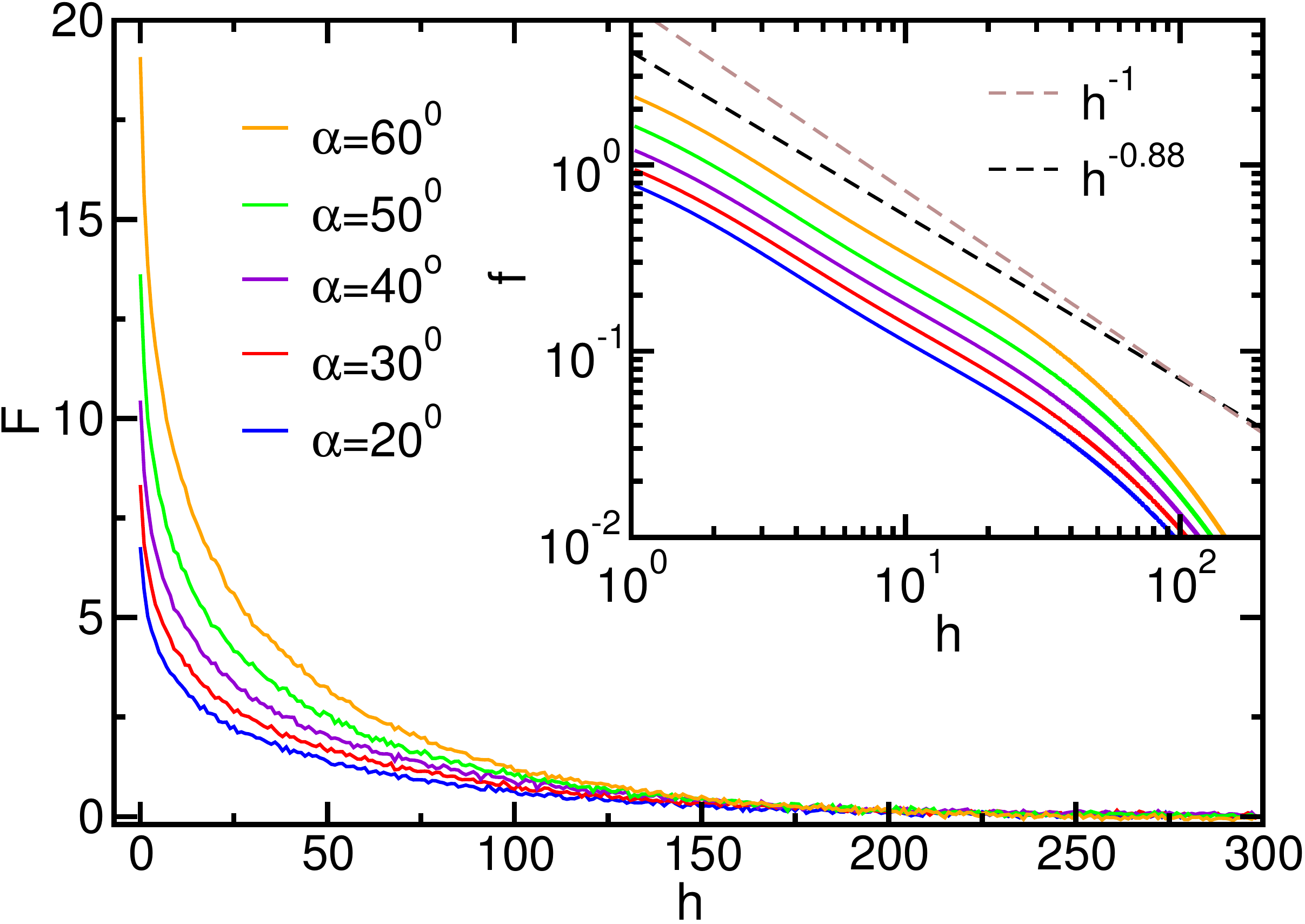}
\end{center}
\caption{Free energy $F$ vs tip-to-surface distance $h$ for a polymer of length
$N$=5000. Results for different cone angles are shown.
The inset shows the variation of the entropic force, $f\equiv |dF/dh|$, with $h$.
The dashed lines show the functions proportional to $h^{-0.88}$ and $h^{-1}$.  }
\label{fig:F.N5000}
\end{figure}

Figure~\ref{fig:F.N5000} shows the variation of $F$ with $h$ for a $N$=5000 polymer  tethered
to a cone with several different values of $\alpha$. As in Fig.~\ref{fig:F.alpha50}, these
results are used to calculate the variation of the entropic force $f$ with $h$. Those results
are shown in the inset of the figure. As in Fig.~\ref{fig:F.alpha50}, there is a range of $h$
over which $f$ varies approximately inversely with $h$. Again, however, the measured exponent
of $-0.88$ is somewhat smaller in magnitude than the predicted value. 
For comparison, a curve with the predicted exponent of $-1$ is overlaid on the graph. As before,
the discrepancy between theory and simulation likely arises from the approximations 
employed in the derivation of Eq.~(\ref{eq:ftheory}). Note that $f(h)$ increases with 
increasing $\alpha$.
This is consistent with the prediction that the scaling prefactor of Eq.~(\ref{eq:ftheory}) is
proportional to $\Delta\gamma$ and the fact that $\Delta\gamma$ increases with $\alpha$, as
illustrated in Fig.~\ref{fig:delF.N}(b).

\begin{figure}[!ht]
\begin{center}
\vspace*{0.2in}
\includegraphics[width=0.45\textwidth]{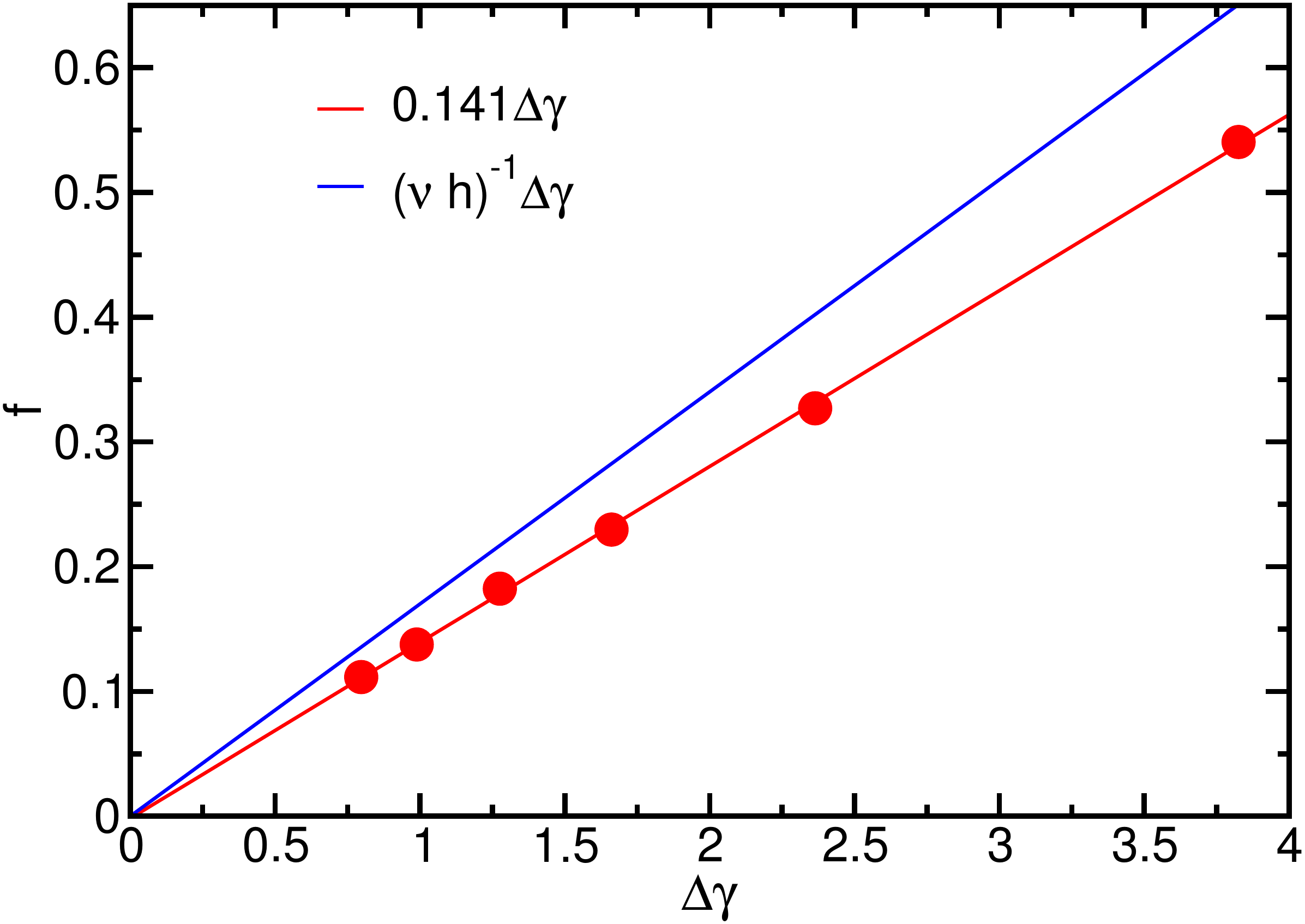}
\end{center}
\caption{Entropic force $f$ vs $\Delta\gamma$ for $N$=5000 and $h$=10 obtained from
the data of the inset of Fig.~\ref{fig:F.N5000}. The red line is a linear fit with a 
slope of 0.141. The blue curve is the prediction using 
Eq.~(\ref{eq:ftheory}) using $h=10$ and $\nu=0.588$. {The uncertainties
in the data points are smaller than the symbol size.}}
\label{fig:fdg}
\end{figure}

Figure~\ref{fig:fdg} shows the variation of $f$ with the $\Delta\gamma$ using the data
from the inset of Fig.~\ref{fig:F.N5000} for a cone tip-to-surface distance value 
of $h=10$, as well as the values of $\Delta \gamma$ obtained from the data of
Fig.~\ref{fig:delF.N}. This value of $h$ lies in the region where $f$ exhibits 
the power-law dependence on $h$ and marginally satisfies the condition 
$a\ll h \ll R_{\rm g}$ required for the prediction of Eq.~(\ref{eq:ftheory}) to be 
valid. Equation~(\ref{eq:ftheory}) predicts that $f$ varies linearly with
$\Delta\gamma$ with a slope of $1/(\nu h)=0.1700$, shown as the blue curve in the 
figure. By   contrast, the simulation data yields a linear dependence, but with a 
slope of $0.141$. Once again, the approximations employed in the derivation of
Eq.~(\ref{eq:ftheory}) lead to small but significant quantitative discrepancies with
the simulation results.

In the limit $\alpha\rightarrow 90^\circ$, the cone becomes a plane, and thus 
the polymer is confined to the space between two parallel surfaces.
This system has been the subject of numerous studies in recent decades, and the dependence
of the conformational free energy on inter-plane spacing and polymer length is well
characterized for both flexible and semi-flexible chains. In the case of flexible
chains in the de~Gennes regime, where $a \ll h\ll R_{\rm g}$ (for monomer size $a$), 
the confinement free energy is equal to the number of thermal blobs, 
which leads to the scaling $F\sim N h^{-1/\nu} \approx N h^{-1.7}$, for $\nu\approx 0.588$. 
This yields an entropic force of $f\equiv |dF/dh| \sim N h^{-1/\nu-1} \approx N h^{-2.7}$. 
The scaling exponent of $-2.7$ is significantly different from the value of $-1$ predicted
from Eq.~(\ref{eq:ftheory}). Note that the fact that the chain is tethered to a point on 
the plane (i.e. the $\alpha=90^\circ$ ``cone'') is only expected to have a tiny effect on 
$F(h)$ and so does not explain the discrepancy.\cite{polson2019free} Evidently, there is a
regime scale  crossover as $\alpha$ increases that is not accounted for in the derivation 
of Eq.~(\ref{eq:ftheory}). 

\begin{figure}[!ht]
\begin{center}
\vspace*{0.2in}
\includegraphics[width=0.4\textwidth]{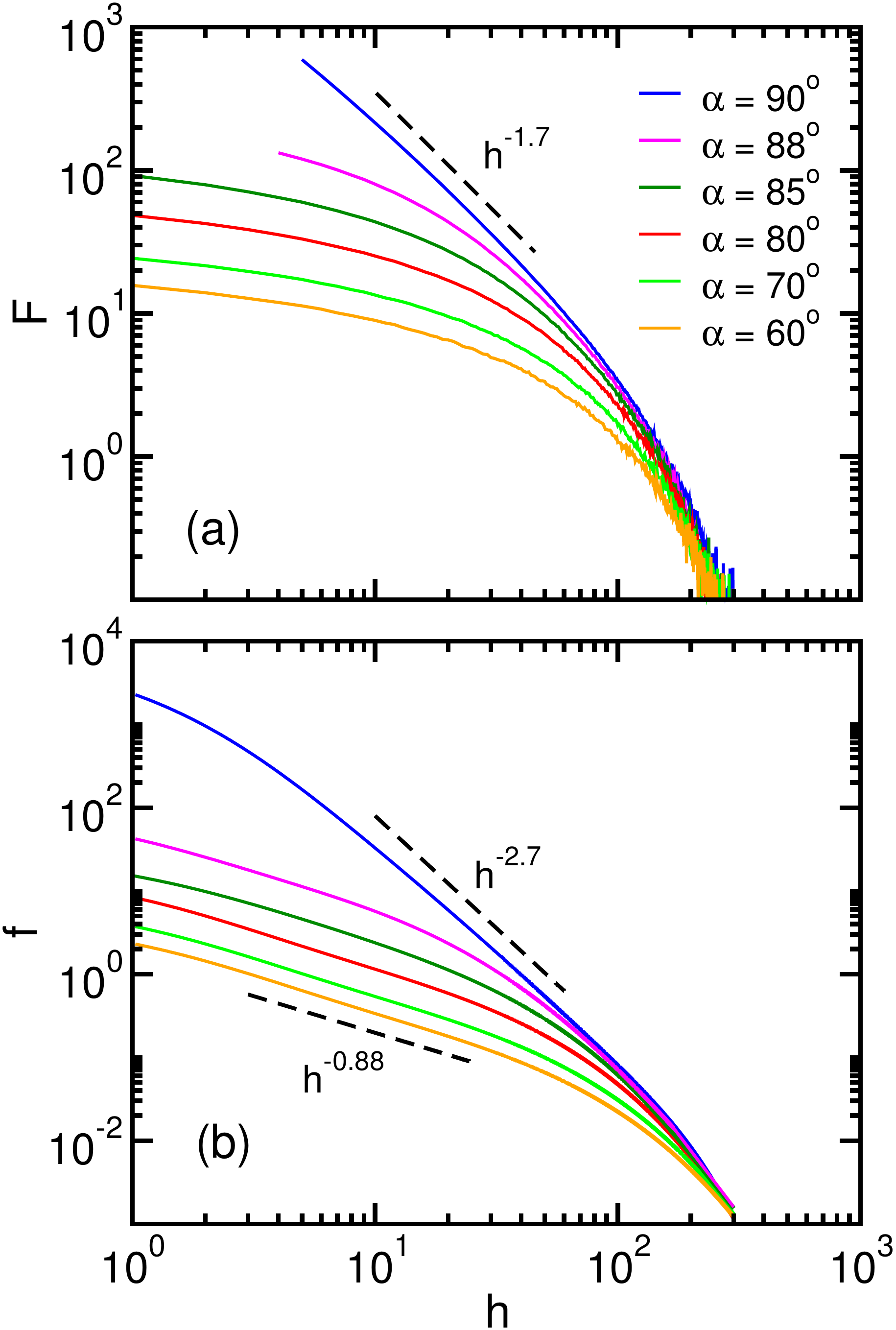}
\end{center}
\caption{(a) Free energy vs cone-tip-to-surface distance $h$ for a polymer of length
$N$=5000 tethered to a cone. Values for various values of $\alpha$ are shown. The dashed
line is the predicted power-law scaling for slit confinement in the de~Gennes regime.
(b) Entropic force calculated using the data of panel (a). The dashed lines show the
prediction for slit confinement in the de~Gennes regime ($f\sim h^{-2.7}$)
and the observed scaling for $\alpha \lesssim 60^\circ$ noted in Fig.~\ref{fig:F.N5000} 
($f\sim h^{-0.88}$).}
\label{fig:F.alpha.wall}
\end{figure}

To elucidate this crossover, we show the variation of $F(h)$ with
cone angle in the range $\alpha\in[60^\circ,90^\circ]$ in Fig.~\ref{fig:F.alpha.wall}(a).
For the case of confinement to a slit ($\alpha=90^\circ$), the de~Gennes-regime scaling
of $F\sim h^{-1.7}$ is observed for intermediate values of $h$. At large $h$, the
power-law exponent increases. The curves for all cone angles converge in the limit
of large $h$. As the distance $h$ decreases, the curves for $\alpha<90^\circ$ diverge
from the $\alpha=90^\circ$ curve. Unsurprisingly, a free energy curve hugs more closely
to the $\alpha=90^\circ$ curve the greater the cone angle. Figure~\ref{fig:F.alpha.wall}(b)
shows the variation of the entropic force $f$ with $h$ calculated using the data of
Fig.~\ref{fig:F.alpha.wall}(a). The crossover between the predicted power-law scaling
for slit confinement in the de~Gennes regime ($f\sim h^{-1/\nu -1}\approx h^{-2.7}$) to
the scaling close to that predicted in Eq.~(\ref{eq:ftheory}) (i.e., $f\sim h^{-0.88}$
rather than $f\sim h^{-1}$) with increasing $\alpha$ is clearly illustrated. Note that
the range of $h$ for which the latter scaling regime holds decreases as the cone becomes
sharper, i.e., as $\alpha$ decreases.

\subsection{A single tethered semi-flexible chain}
\label{subsec:semiflexible}

\begin{figure}[!ht]
\begin{center}
\vspace*{0.2in}
\includegraphics[width=0.45\textwidth]{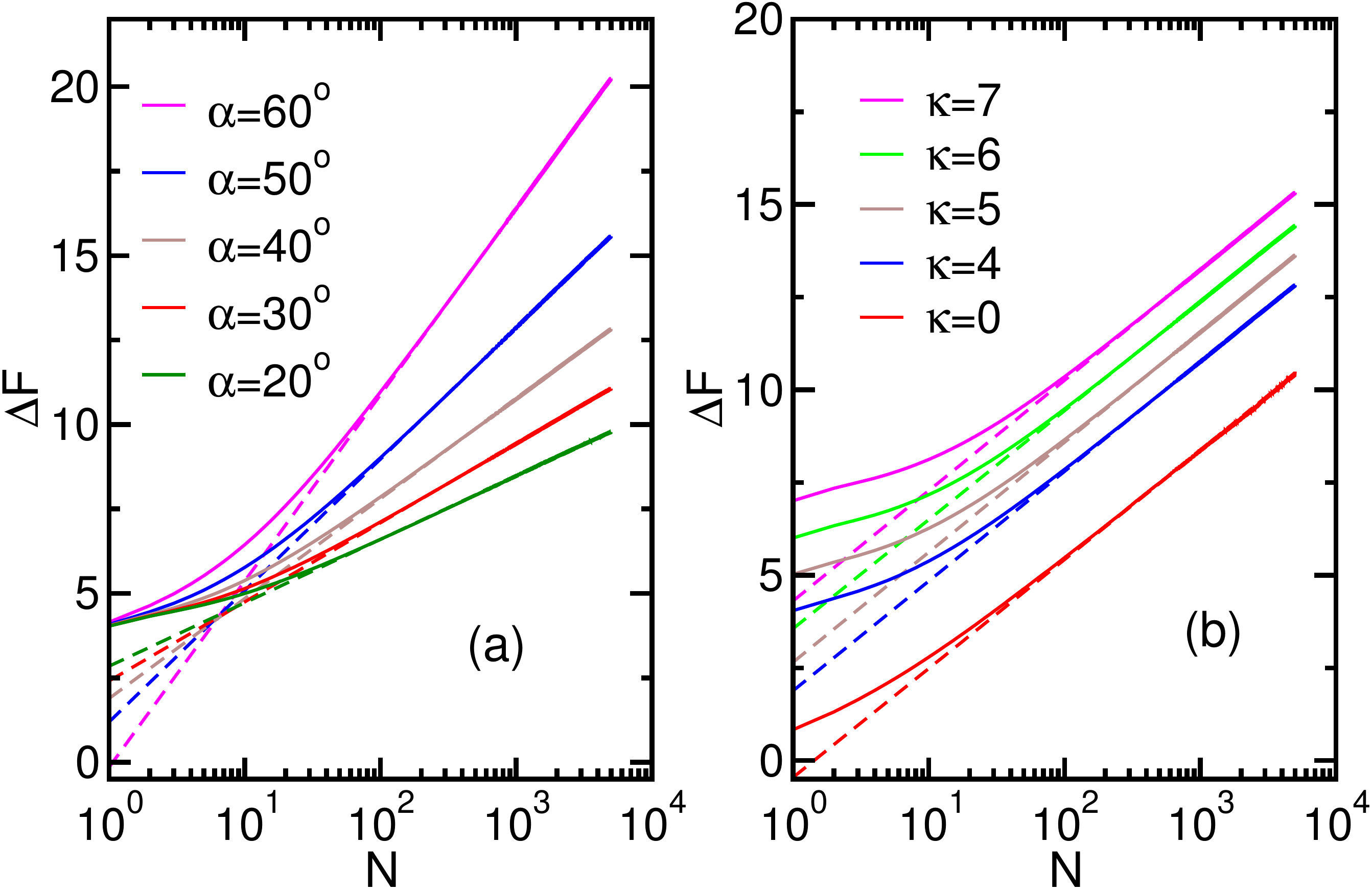}
\end{center}
\caption{(a) Free energy difference $\Delta F \equiv F(h=0)-F(h=\infty)$ vs polymer
length $N$ for a single semiflexible chain  with bending rigidity $\kappa=4$
tethered to a cone of half-angle $\alpha$. Results for various values of $\alpha$ are shown. 
The dashed lines show fits of the data in the region $N\in [300,2000]$ to the function
$\Delta F = a_0 + a_1\ln N$. (b) As in panel (a), except for fixed cone angle of
$\alpha=40^\circ$ and for various values of $\kappa$. }
\label{fig:delF.N.kappa}
\end{figure}

Now we consider the effects of polymer bending rigidity on the entropic force. 
Figure \ref{fig:delF.N.kappa} shows the variation of $\Delta F = F(h=0)-F(h=\infty)$
with $N$ for a chain with a bending rigidity of $\kappa=4$ tethered to a cone for
various values of $\alpha$. As in Fig.~\ref{fig:delF.N}(a), the dashed lines show
fits to the function $F = a_0 + a_1\ln(N)$, in this case for $N>300$, with the
fitted line extended to lower $N$ to highlight the expected discrepancy in this
regime. The value of $a_1$ is an estimate of the proportionality factor $\Delta\gamma$
appearing in Eq.~(\ref{eq:FCDg}). As expected, the values of $\Delta\gamma$ are
unaffected to the introduction of bending rigidity as they are essentially equal to 
the values obtained using freely-jointed chains. This is also illustrated clearly
for the case of $\kappa$=4, for which the values of $\Delta\gamma$ vs $\alpha$ 
are overlaid on those for flexible chains in Fig.~\ref{fig:delF.N}(b).
The curves for $\Delta F$ vs $N$ for the semiflexible chains differ from those for 
freely-jointed chains in Fig.~\ref{fig:delF.N}(a) only by a shift of the curves to higher free 
energy. This arises because of the  energy required to bend the polymer near the cone tip 
in the case of $h=0$, where surface is brought in contact with the cone tip. This shift 
in $\Delta F$ is expected to increase with increasing $\kappa$.  Note that 
$\Delta F\approx \kappa$ for $N$=1 for the $\kappa>0$ curves. This is because the free 
energy change is dominated by the change in the bending energy associated with the 
monomer connected to the one fixed at the cone tip. Placing the flat surface at $h$=0
forces the monomer to bend at least  $90^\circ$, with greater angles discouraged 
by a still higher bending energy. At this angle, the bending energy is 
$u_{\rm bend}=\kappa (1-\cos(90^\circ))=\kappa$.

\begin{figure}[!ht]
\begin{center}
\vspace*{0.2in}
\includegraphics[width=0.45\textwidth]{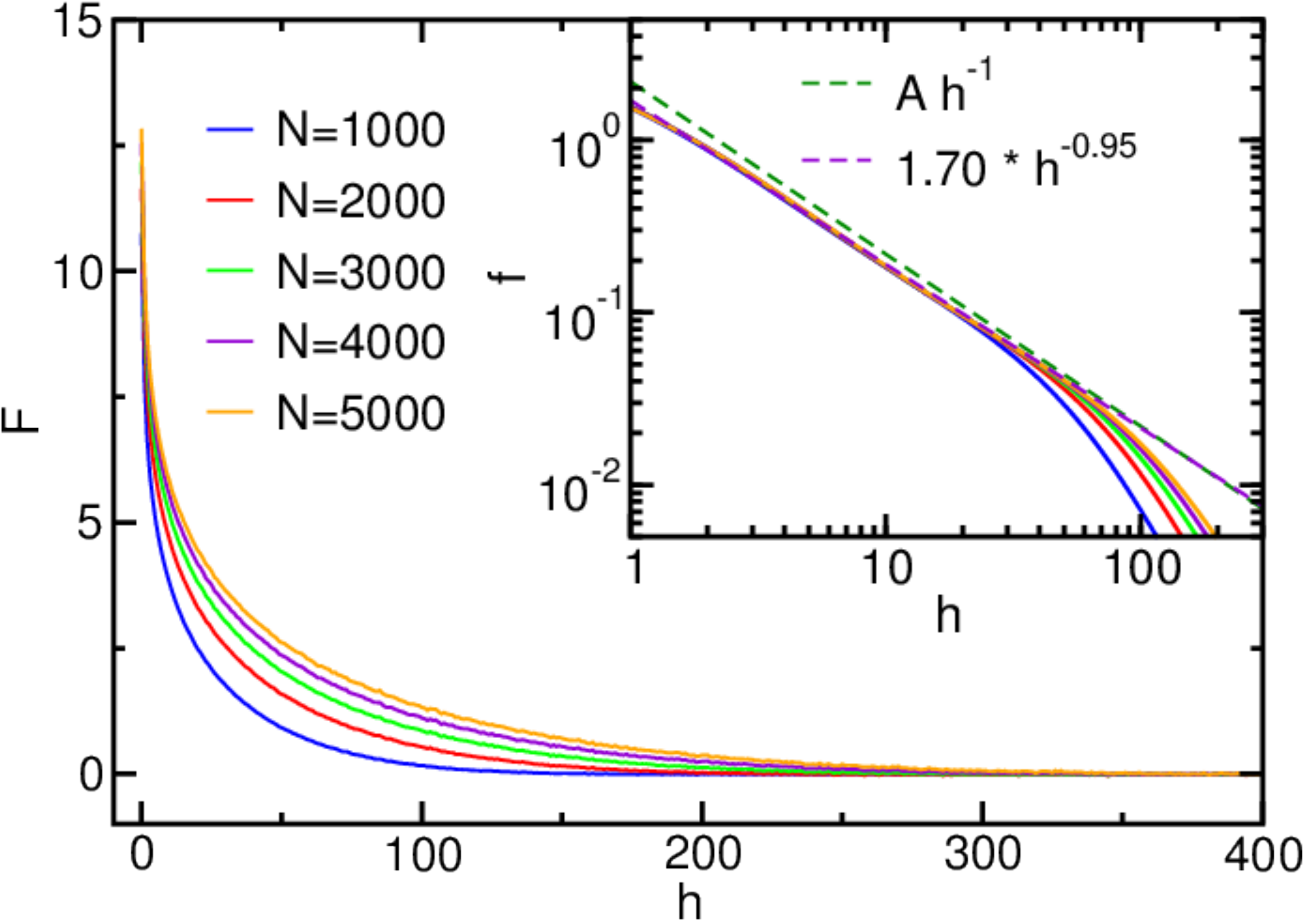}
\end{center}
\caption{ Free energy $F$ of a semiflexible tethered chain vs tip-to-surface distance $h$ 
for a cone angle of $\alpha=40^\circ$. Results for various polymer lengths are shown.
The inset shows the variation of the entropic force, $f\equiv |dF/dh|$, with $h$.
The green dashed line shows the theoretical prediction using the value of
$\Delta\gamma$ extracted from the fit in Fig.~\ref{fig:delF.N.kappa}. The dotted
black line shows a fit of the $N=5000$ curve to the function $f=cN^\mu$
in the region $h\in [4,30]$.  }
\label{fig:F.alpha40.kappa4}
\end{figure}

Figure~\ref{fig:F.alpha40.kappa4} shows the variation of the free energy with $h$ 
for a semiflexible chain with $\kappa=4$ tethered to a cone with $\alpha=40^\circ$. 
Results for various chain lengths are shown. The inset shows the variation of
the corresponding entropic force, $f$, with $h$. As for the behavior $f(h)$ for freely 
jointed chains
shown in the inset of Fig.~\ref{fig:F.alpha50}, the curves for different $N$ collapse
on a single curve for a restricted range of $h$, which decays with a power law. 
As $h$ increases further, each curve eventually peals away from the others in the
order from shortest to longest. The green dashed curve shows the prediction of 
Eq.~(\ref{eq:ftheory}), which is a closer match to the simulation data than was the
case for the freely-jointed chains. {The scaling exponent of the fit to the $N=5000$
curve in the range $h\in[4,30]$ is $-0.95\pm 0.03$}, which is close to 
the predicted value of $-1$. 

\begin{figure}[!ht]
\begin{center}
\vspace*{0.2in}
\includegraphics[width=0.45\textwidth]{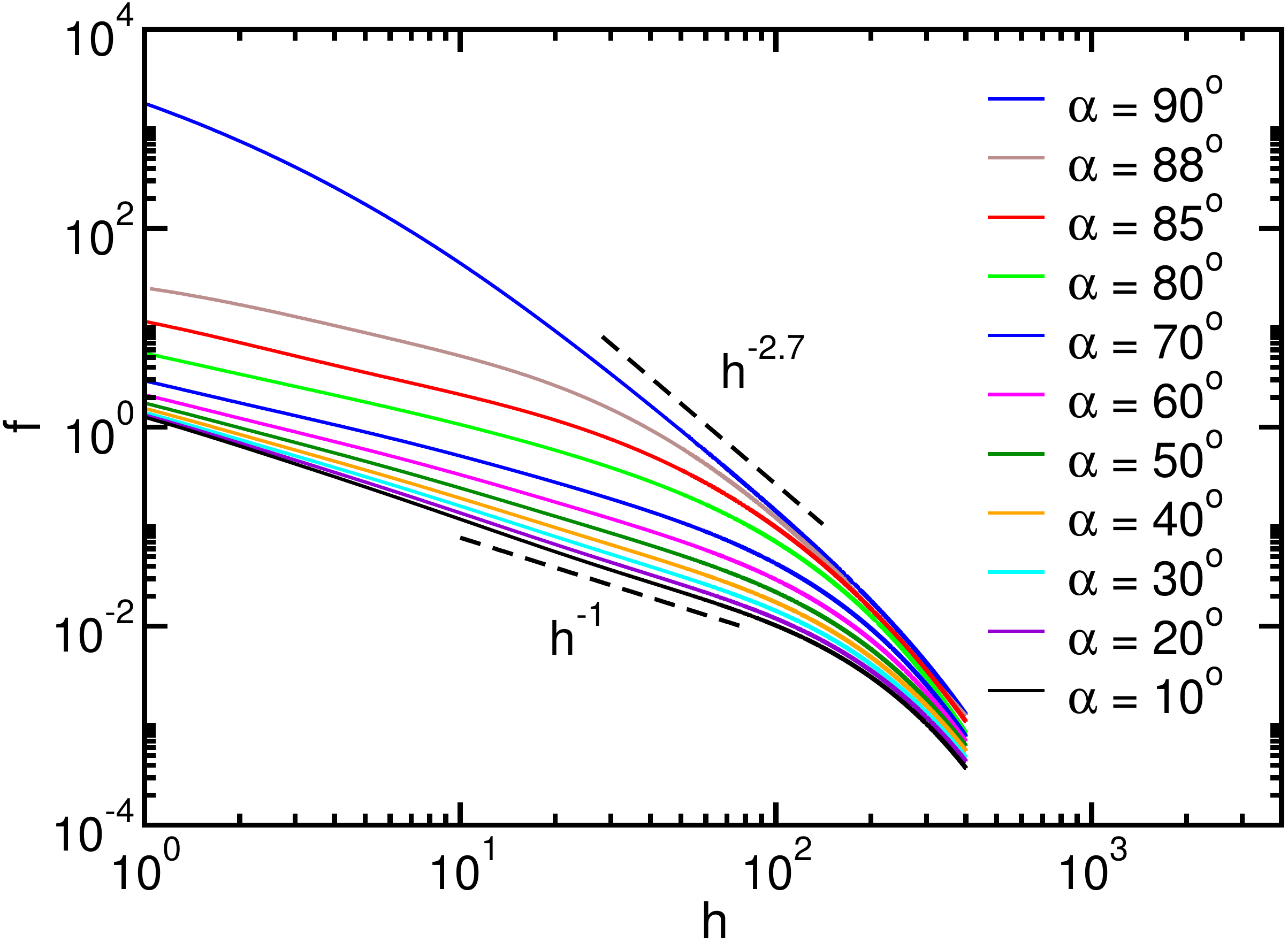}
\end{center}
\caption{Entropic force $f$ vs tip-to-surface distance $h$ for a polymer of length
$N$=5000 and bending rigidity $\kappa=4$. Results for different cone angles are shown.
The dashed lines show the prediction of Eq.~(\ref{eq:ftheory}) ($f\sim h^{-1}$) as well
as the prediction for slit confinement in the de~Gennes regime ($f\sim h^{-2.7} \approx h^{-1/\nu-1}$).}
\label{fig:f.alpha.k4}
\end{figure}

Figure~\ref{fig:f.alpha.k4} shows the variation of the entropic force, $f$, with respect
to $h$ for a number of cone angles in the range $\alpha \in [10^\circ,90^\circ$]. 
As was the case for flexible chains in Fig.~\ref{fig:F.alpha.wall}(b), the functions
exhibit power-law behavior for an intermediate range of $h$. In the case of slit
confinement ($\alpha=90^\circ$), the scaling exponent in this regime is consistent with
the predictions for the de~Gennes regime of $f\sim h^{-1/\nu-1} \approx h^{-2.7}$ for
distances roughly in the range $h\in [20,200]$. (Note that the extended de~Gennes
regime for slit confinement is present only if the condition $2P < h < 0.2P^2/w$
is satisfied, where $P$ is the persistence length and $w$ is the polymer 
width.\cite{cheong2018evidence} Noting that $\kappa=4$ yields $P\approx 4$ and 
also that $w$=1, the extended de~Gennes is not expected to be present for this 
system.) As the cone angle decreases, the scaling approaches the prediction of 
Eq.~(\ref{eq:ftheory}) of $f\sim h^{-1}$. The change in the curves as $\alpha$ 
increases from $10^\circ$ to $90^\circ$ is somewhat more complex than the trends 
for flexible chains evident in the inset of Fig.~\ref{fig:F.N5000} and in 
Fig.~\ref{fig:F.alpha.wall}(b). 

\begin{figure}[!ht]
\begin{center}
\vspace*{0.2in}
\includegraphics[width=0.45\textwidth]{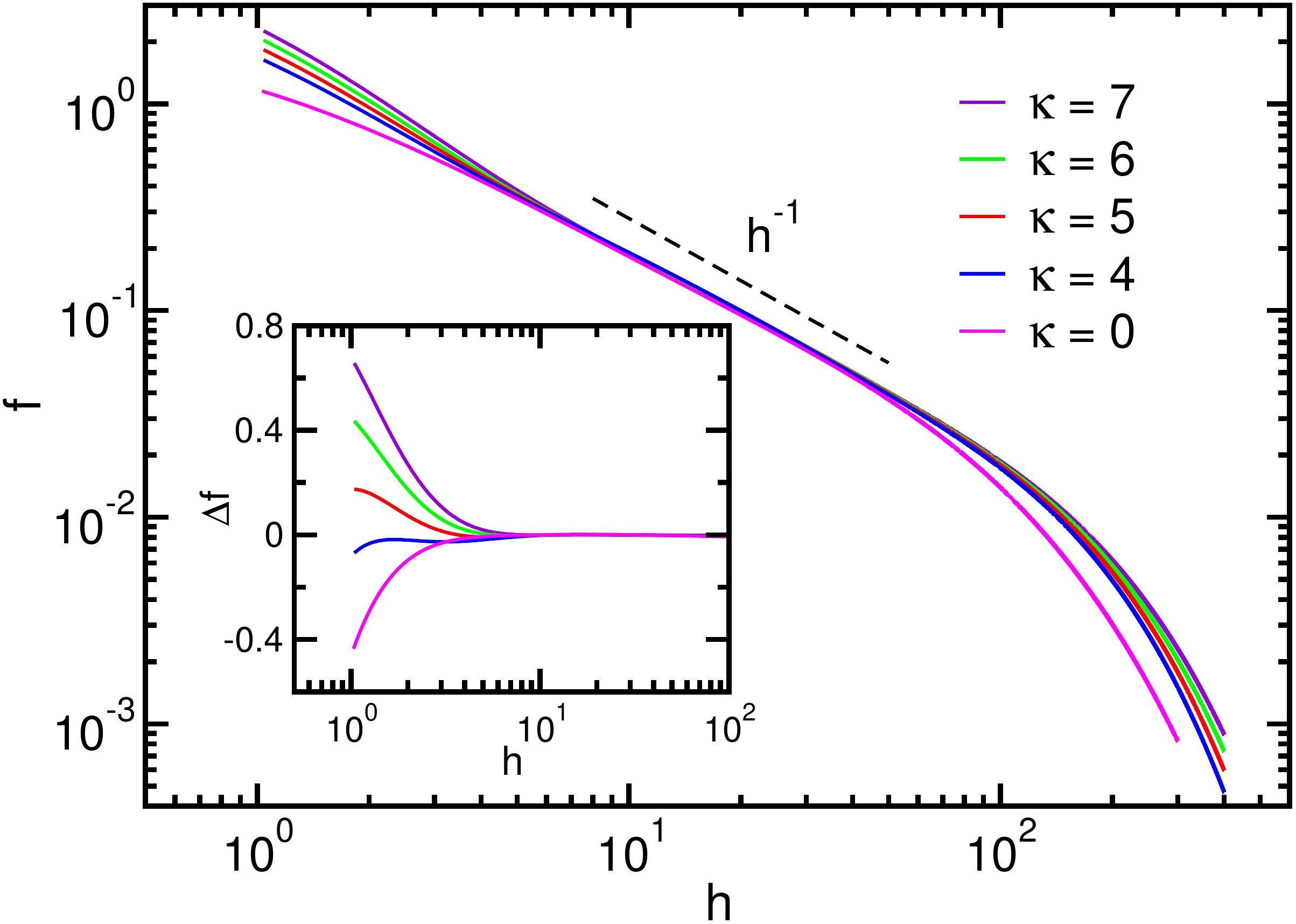}
\end{center}
\caption{Entropic force $f$ vs tip-to-surface distance $h$ for a polymer of length $N$=5000 
and cone angle of $\alpha=40^\circ$. Results for various values of the bending rigidity are 
shown. The inset shows the difference $\Delta f = f - f^*$ vs $h$, where $f^*(h)$ is the 
fit of each curve to the function $f^* = c_0 h^{\mu}$ in the range $h\in[8,30]$.}
\label{fig:f.N5000.alpha40.kappa}
\end{figure}

Figure~\ref{fig:f.N5000.alpha40.kappa} shows force-distance functions for a polymer
of length $N$=5000 tethered to a cone of angle $\alpha=40^\circ$ for various values
of the bending rigidity, $\kappa$. Consistent with the prediction of Eq.~(\ref{eq:ftheory}),
$f$ is independent of molecular details such as the bending rigidity over an intermediate
range of $h$. At lower and higher values of $h$, however, the curves diverge slightly. 
In the case of large $h$, the force, while very weak, does increase somewhat
with increasing $\kappa$. This occurs since stiffer chains are expected to stretch somewhat
further away from the cone and thus interact more significantly with a distant planar 
surface than will more flexible
chains. At shorter distances, a similar trend occurs, but for very different reasons.
The effects are highlighted in the inset of the figure, which shows the $h$-dependence
of the difference, $\Delta f \equiv f - f^*$, where $f^*$ is a fit of $f(h)$ to a
power-law function in the intermediate regime of $h$ where this scaling holds. 
At low $h$, $\Delta f$ deviates significantly from zero. In the case of a freely-jointed 
chain ($\kappa=0$), $\Delta f$ is negative and decreases as $h$ decreases. 
This deviation arises from violation of the condition that $a\ll h$ for
Eq.~(\ref{eq:ftheory}) to be valid. Upon increasing $\kappa$, $\Delta f$ also increases,
becoming positive for $\kappa\gtrsim 4$. The physical origin of this trend is 
straightforward. A stiff polymer tethered to a cone tip close to the flat surface  
is forced to bend, giving rise to an appreciable bending energy and corresponding 
elastic force that pushes on the surface. This elastic force naturally increases
with $\kappa$, giving rise to the observed trend at small $h$.

\subsection{Multiple tethered freely-jointed chains}
\label{subsec:multiple}

\begin{figure}[!ht]
\begin{center}
\vspace*{0.2in}
\includegraphics[width=0.45\textwidth]{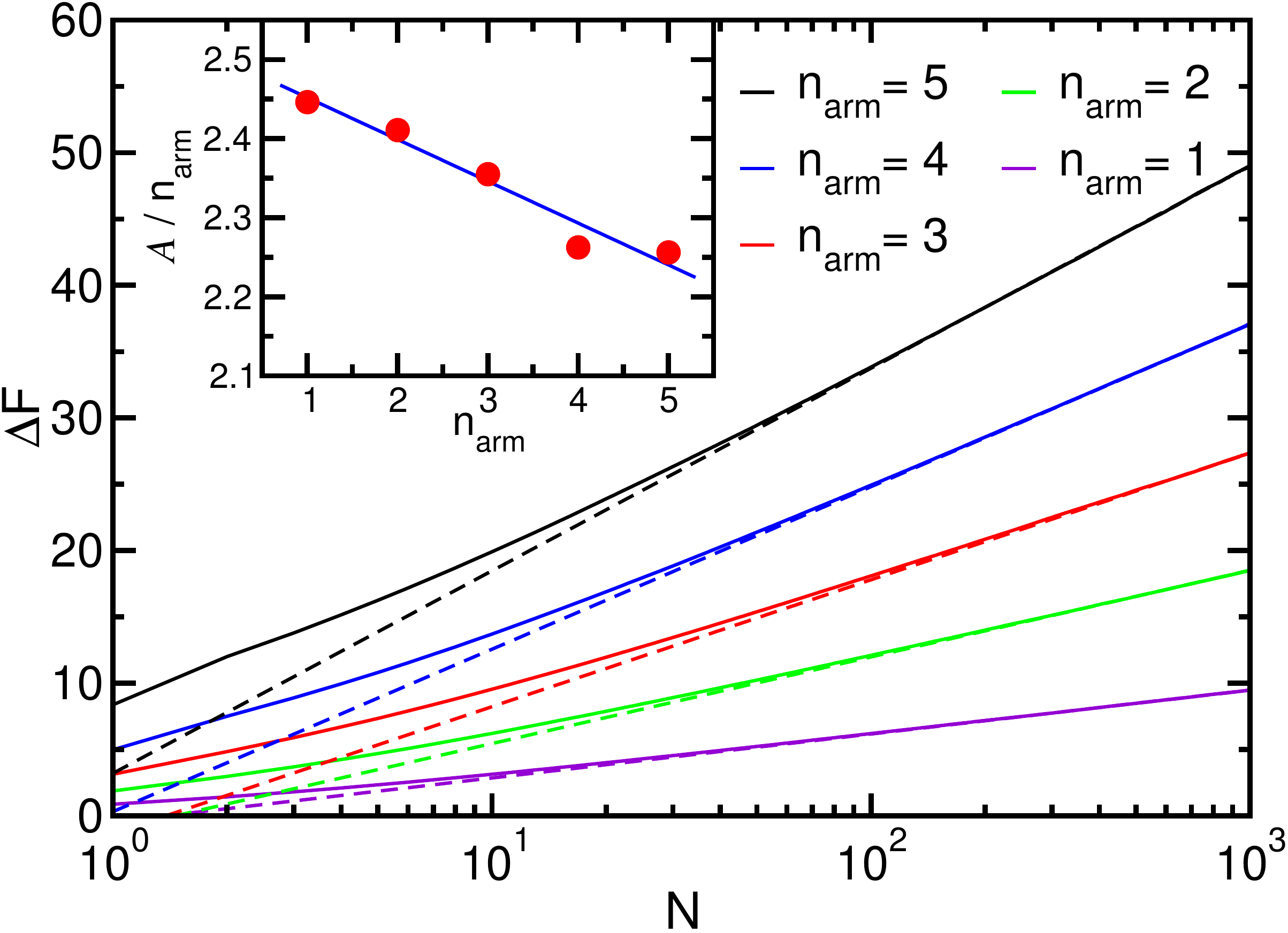}
\end{center}
\caption{
Free energy difference, $\Delta F\equiv F(h=0)-F(h=\infty)$, for a star polymer
with $n_{\rm arm}$ arms and with the branch point tethered to the tip of the cone. 
Results are show for a cone angle of $\alpha=45^\circ$  and a star polymer with the
length of each arm up to $N=1000$. Results are shown for $n_{\rm arm}=$1--5. 
The inset shows the variation of ${\cal A}/n_{\rm arm}$ vs $n_{\rm arm}$, where 
${\cal A}\equiv\Delta\gamma/\nu$ and where $\Delta \gamma$ is obtained from the fit 
$\Delta F = {\rm const} + \Delta \gamma N$ in the region $N\in[600,1000]$.
{The blue line is a linear fit to the data.} }
\label{fig:delF.N1000.narm}
\end{figure}

We now consider the case of multiple polymers, each of length $N$ and each tethered to the
tip of the cone. This can also be viewed as a single star polymer with the branch point
fixed to the cone tip, where the length of each arm is $N$. We denote the number of
arms of the star polymer as $n_{\rm arm}$. Figure~\ref{fig:delF.N1000.narm} shows the 
variation of $\Delta F\equiv F(h=0)-F(h=\infty)$ with $N$. Results for a cone of angle 
$\alpha=45^\circ$ and   
for a few different values of $n_{\rm arm}$ are shown. As in Figs.~\ref{fig:delF.N} and
\ref{fig:delF.N.kappa}, $\Delta F$ varies with $N$ as $F=a_0 + a_1\ln N$ at sufficiently
high $N$, consistent with the form of Eq.~(\ref{eq:FCDg}). Fitting each data set to
this function yields an estimate for $\Delta \gamma$ from the fitting parameter $a_1$. 
In Refs.~\onlinecite{maghrebi2011entropic} and \onlinecite{maghrebi2012polymer},
Maghrebi {\it et al.} used the $\epsilon$-expansion to predict that
\begin{eqnarray}
\frac{\cal A}{n_{\rm arm}} = 1 - \frac{\epsilon}{8} 
+\left[ \frac{3}{\pi} -\left( 0.80 + \frac{11}{12\pi}(n_{\rm arm} -1)\right)\epsilon \right]
\alpha^{1-3\epsilon/4}, \nonumber \\
\end{eqnarray}
where ${\cal A}\equiv \Delta\gamma/\nu$ is the prefactor appearing in the force-distance 
relation of Eq.~(\ref{eq:fdef}), and where $d-3=1-\epsilon$. Using $d=3$ dimensions, 
and $\alpha=45^\circ = \pi/4$~rad for the cone used in these calculations, it follows that 
\begin{eqnarray}
\frac{\cal A}{n_{\rm arm}} \approx {\rm const.} - {\cal B}n_{\rm arm},
\label{eq:Adivnarm}
\end{eqnarray}
where ${\cal B}=(11/12\pi)\alpha^{1/4}$. Thus, the theory predicts that the entropic force
per arm {\it decreases} as the number of arms increases. For the cone used in these
calculations, $\alpha=45^\circ = \pi/4$~rad, yielding a value of ${\cal B}=0.275$. 
To test this prediction, we plot ${\cal A}/n_{\rm arm}$ vs $n_{\rm arm}$ in the inset
of the figure, where
${\cal A}\equiv \Delta \gamma/\nu$, and where $\Delta\gamma$ are obtained from the fits
described above. The results are shown in the inset of the figure. Consistent with the
theoretical prediction, we find that ${\cal A}/n_{\rm arm}$ does appear to decrease with
increasing $n_{\rm arm}$. However, {the slope of the best linear fit is $0.053\pm 0.007$},
which is about a factor of 5.2 smaller than the prediction for ${\cal B}$. For the case of cone
angle $\alpha=45^\circ$, the predicted force per arm decreases by roughly 10\%.

\begin{figure}[!ht]
\begin{center}
\vspace*{0.2in}
\includegraphics[width=0.45\textwidth]{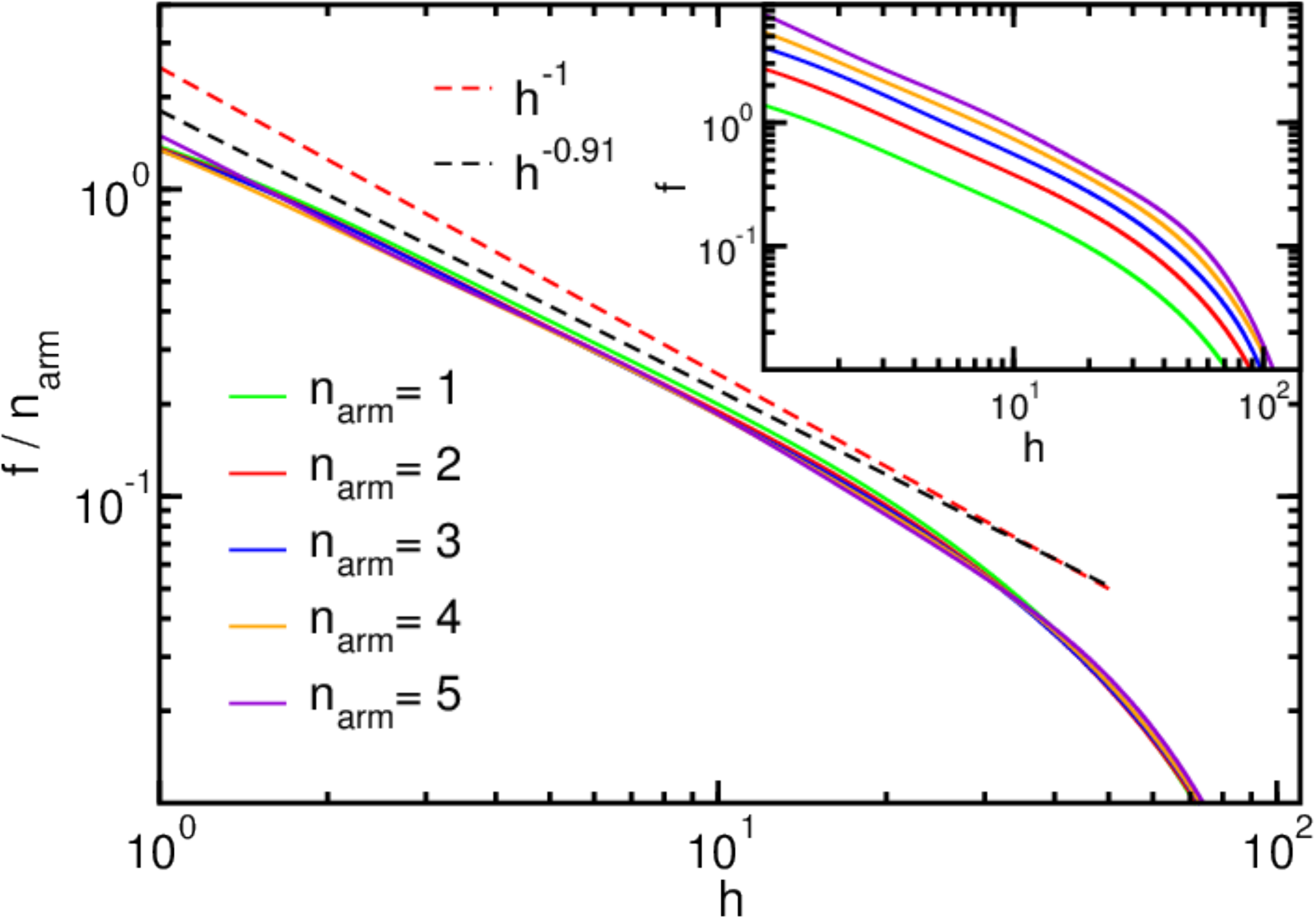}
\end{center}
\caption{Entropic force $f$ per arm vs tip-to-surface distance $h$ for a star polymer 
with $n_{\rm arm}$ arms and with the branch point tethered to the tip of the cone.
Results are shown for an arm length $N$=1000 and cone angle of $\alpha=45^\circ$. 
Results for various values of $n_{\rm arm}$ are shown. The inset shows $f$ vs $h$ for the
same data sets.}
\label{fig:f.N1000.narm.alpha45}
\end{figure}

Figure~\ref{fig:f.N1000.narm.alpha45} shows the variation in the measured entropic force 
per arm with $h$ for a star polymer of arm length $N=1000$ with the branch point fixed to 
the  tip of a cone of angle $\alpha=45^\circ$. Results for a few values of $n_{\rm arm}$ 
are shown. The inset shows $f$ vs $h$ for the same data sets. As was the case for single
tethered polymers, there is an intermediate range of $h$ over which the relation
$f={\cal A}/h$ is approximately satisfied. {A fit to each of the data sets in the range 
$h\in [3,13]$  yields estimated exponents of $-0.92\pm 0.04$, $-0.94\pm 0.04$, $-0.90\pm 0.02$, 
$-0.91\pm 0.02$, and $-0.95\pm 0.06$, for $n_{\rm arm}$= 1, 2, 3, 4 and 5, respectively.} 
Thus, as in previous cases, the scaling exponent differs slightly from the predicted value 
of $-1$.  Contrary to the the prediction of Eq.~(\ref{eq:Adivnarm}) and the trend observed 
in Fig.~\ref{fig:delF.N1000.narm}, there is no clear evidence that the entropic force
per arm decreases with $n_{\rm arm}$. While it is possible that such an effect is masked 
by  statistical limitations of the data, its magnitude is unlikely to be anywhere near
to that predicted by the epsilon-expansion. Force-distance data generated from simulations
using much longer chain lengths would be helpful to better elucidate this subtle effect,
though this is not currently feasible. For the present, we tentatively conclude that
total entropic force is simply proportional to the number of arms of a cone-tethered star
polymer to an excellent approximation.

\section{Conclusions}
\label{sec:conclusions}

In this study, we have used PERM MC simulations to characterize the entropic force
between a hard conical surface and a hard planar surface mediated by one or more
polymers tethered to the tip of the cone. These calculations were inspired by previous
work by Maghrebi {\it et al.},\cite{maghrebi2011entropic, maghrebi2012polymer} who
predicted that the force should obey a scaling relation $f={\cal A} k_{\rm B}T/h$,
where $h$ is the cone-tip-to-surface distance. The relation is expected to hold
for sufficiently long polymers in the regime $a\ll h \ll R_{\rm g}$, where $a$ is
the polymer segment length and $R_{\rm g}$ is the radius of gyration for a free polymer.
The prefactor is given by  ${\cal A} = (\gamma_\infty-\gamma_0)/\nu$, where $\gamma_\infty$
and $\gamma_0$ are critical exponents appearing in the partition function of the tethered
polymer(s) and where $\nu$ is the the Flory exponent for a self-avoiding polymer. We 
measured the force-distance relation for a single fully-flexible and semi-flexible 
cone-tethered hard-sphere chain, as well as a system with multiple polymers end-tethered
to the cone. In each case, we find that there is indeed an intermediate range of $h$ for
which the proportionality $f\propto h^{-1}$ approximately holds, though the scaling 
exponent tends to be somewhat smaller than the predicted value of 1. Simulations for $h=0$
and $h=\infty$ facilitated calculation of the critical exponents $\gamma_0$ and
$\gamma_\infty$ and, thus, of the scaling prefactor ${\cal A}$. As in the case of the 
scaling exponent, we find the measured value of ${\cal A}$ to be slightly smaller than 
the predicted value. These small discrepancies presumably arise from the various
approximations employed in the analytical theory. On the other hand, within the valid 
range of $h$, the entropic force is independent of chain length and chain stiffness, and
the scaling of the force with $h$ is independent of the number of
polymers tethered to the cone, all in accord with the prediction. In the case of
multiple tethered polymers, we find that the entropic force scales proportional to the
number of polymers, at least to within the precision of the calculations. This is in
disagreement with the predictions from $\epsilon$-expansion calculations carried out
by Maghrebi {\it et al.}, in which the force per polymer is expected to {\it decrease} with
the number of tethered polymers. Thus, this analytical method significantly overestimates
any such effect. 

As noted in Sec.~\ref{sec:intro}, the theoretical prediction of Maghrebi {\it et al.}
has been tested in a recent AFM experiment by Liu {\it et al.}\cite{liu2019measurement}
Polyethylene glycol
(PEG) polymers were tethered to a pyramidal AFM tip, and the force between the tip and
a flat surface were measured as a function of the tip-to-surface distance. To measure
the entropic force the van~der~Waals forces were eliminated by subtraction of the force
measured in a separate experiment carried out with no polymers attached. Conditions
were tuned to eliminate electrostatic forces and polymer adhesion to the surface. 
In spite of these efforts, the theory provided a poor prediction for the observed trends,
proving inferior to the predictions using the Alexander-de~Gennes (AdG) theory for a 
confined polymer brush.\cite{deGennes_book} An
obvious source of the discrepancy arises from the structure of the AFM tip employed.
In order to end-tether the PEG polymers, they were covalently bound to a well-defined
Au patch at the apex of the cantilever tip, which had a surface area of about 
$3.7\times 10^4$~nm$^2$. On the other hand, the radius of gyration of the PEG molecules
was only about 14~nm and 22~nm for the two different PEG molecular weights considered.
Given these length scales, the system  better resembles slit-confined surface-tethered
polymers rather than the cone-plane-confined system described in
Refs.~\onlinecite{maghrebi2011entropic} and \onlinecite{maghrebi2012polymer} as well as
the present study, thus accounting for the better agreement with the AdG theory. 
A better experimental test of the theory would require using much longer polymers or
reducing the size of the patch at the cantilever tip apex. Even so, the truncation 
of the AFM tip to produce the patch introduces a new length scale that is expected to
complicate the prediction that $f={\cal A}k_{\rm B}T/h$. In future work,
we will use the computational methods employed here to quantify this effect. 
Another relevant feature to incorporate into the model is roughness of the planar surface
(to which Liu {\it et al.} attribute anomalous behavior of the force at very low tip-to-surface
distances). Measurements of the entropic force for such systems should be helpful for
optimizing the design of future experiments to better test the theoretical prediction. 

\begin{acknowledgements}
This work was supported by the Natural Sciences and Engineering Research Council of Canada
(NSERC) Discovery Grants Program.  We are grateful to Compute Canada for use of their
computational resources.
\end{acknowledgements}


%

\end{document}